\documentclass[aps,prd,superscriptaddress,amsmath,twocolumn,10pt]{revtex4}

\usepackage{color,hangcaption,hhline,psfrag,rotating,amssymb}
\usepackage[hang,nooneline]{subfigure}
\usepackage{dcolumn}
\usepackage{verbatim}
\usepackage{multirow}

\newcommand{\nn}{\nonumber}
\newcommand{\be}{\begin{equation}}
\newcommand{\ee}{\end{equation}}
\newcommand{\bea}{\begin{eqnarray}}
\newcommand{\eea}{\end{eqnarray}}

\newcommand{\order}{\mathcal{O}}

\begin{document}

\title{Bottomonium hyperfine splittings from lattice NRQCD including radiative and relativistic corrections}

\author{R.~J.~Dowdall}
\affiliation{DAMTP, University of Cambridge, Wilberforce Road, Cambridge CB3 0WA, UK}
\author{C.~T.~H.~Davies}
\email[]{christine.davies@glasgow.ac.uk}
\affiliation{SUPA, School of Physics and Astronomy, University of Glasgow, Glasgow, G12 8QQ, UK}
\author{T.~Hammant}
\author{R.~R.~Horgan}
\author{C. Hughes}
\affiliation{DAMTP, University of Cambridge, Wilberforce Road, Cambridge CB3 0WA, UK}

\collaboration{HPQCD collaboration}
\noaffiliation
\homepage{http://www.physics.gla.ac.uk/HPQCD}

\date{\today}

\begin{abstract}
We present a calculation of the hyperfine splittings in bottomonium using lattice Nonrelativistic QCD.
The calculation includes spin-dependent relativistic corrections through $\order(v^6)$, radiative corrections to the leading spin-magnetic coupling and, for the first time, non-perturbative four-quark interactions which enter at $\alpha_s^2 v^3$.
We also include the effect of $u,d,s$ and $c$ quark vacuum polarisation.
Our result for the 1S hyperfine splitting is $M_\Upsilon(1S) - M_{\eta_b}(1S)= 60.0(6.4)$ MeV. We find the ratio of 2S to 1S hyperfine splittings $(M_\Upsilon(2S) - M_{\eta_b}(2S))/ (M_\Upsilon(1S) - M_{\eta_b}(1S)) = 0.445(28)$.
\end{abstract}


\maketitle

%
\section{Introduction}
\label{sec:intro}
%
The energy splittings between the vector and pseudoscalar bottomonium states have, until recently, proved difficult for both experimentalists and theorists. Experimentally, the $\eta_b(1S)$ was only discovered in $\Upsilon$ radiative decays in 2008 by Babar \cite{Aubert:2008ba}, and subsequent results have led to a fairly wide error being placed on the PDG average \cite{pdg,Aubert:2009pz,Bonvicini:2009hs,belleetab}. The $2S$ state has recently been observed by two groups, Belle \cite{belleetab} and Dobbs et al. \cite{cleoetab2s} using CLEO data, but the results differ signifcantly. Belle finds a $2S$ hyperfine splitting of $24.3^{+4.0}_{-4.5}$ MeV and Dobbs et al. have $48.7\pm3.1$ MeV.
These difficulties are in contrast to the vector particle $\Upsilon$ which was found in 1977 \cite{Herb:1977ek} and whose mass is now known to sub-MeV precision.

On the theory side, predictions of the hyperfine splittings generally suffer from large systematic errors. Potential model estimates quoted in the Quarkonium Working Group review \cite{Brambilla:2004wf} range from 46-87 MeV for the $1S$ splitting \cite{Godfrey:1985xj,Fulcher:1991dm,Ebert:2002pp,Gupta:1995ps,Zeng:1994vj,Eichten:1994gt}, with systematic errors that are hard to quantify. 
Next-to-leading order continuum pertubation theory gives 41(14) MeV \cite{Kniehl:2003ap}.
First principle lattice QCD calculations offer 
the prospect of much more accurate results but 
there too control of systematic errors (as opposed to 
statistical errors) is the key issue. 
The most successful approaches to date have used 
an effective field theory for the b quark such 
as nonrelativistic QCD (NRQCD). The hyperfine splitting 
is generated by an operator in the Lagrangian that 
appears first at $\order(v^4)$ in a power-counting 
in terms of the velocity of the $b$-quark. Including 
only the leading term, as was done in the first calculations, gives the penalty of a large systematic 
error from missing radiative corrections to the leading term and missing higher order terms. 
These systematic errors would be avoided by treating 
the b quark as a relativistic quark but in that 
case significant discretisation errors appear instead. 
It is now becoming possible to handle $b$ quarks relativistically with the Highly Improved Staggered 
Quark action on very fine lattices \cite{McNeile:2011ng, McNeile:2010ji} but results for the 
bottomonium hyperfine splitting are still in the future. 
A middle ground between nonrelativistic and relativistic 
approaches is the Fermilab approach to heavy quarks which 
interpolates between heavy and light regimes. Early work 
yielded rather inaccurate results \cite{Burch:2009az} but improvements 
currently underway for charmonium splittings \cite{DeTar:2012xk} will 
be applied to bottomonium in future. 

Motivated by the experimental discrepancies, we revisit the hyperfine splittings with NRQCD and 
 make the necessary improvements to the action to eliminate the dominant systematic errors. 
In NRQCD the hyperfine splitting arises from the $\mathcal{O}(v^4)$ spin-magnetic coupling term $\sigma.{\bf B}/2m_b$, whose coefficient we denote $c_4$.
 At tree level this action gave a hyperfine splitting of 61(14) MeV \cite{Gray:2005ur} where the large systematic error comes from unknown radiative corrections to $c_4$. 
HPQCD calculated these corrections to one-loop in \cite{Dowdall:2011wh,Hammant:2013sca}. 
Including the corrections along with an estimate of the effect of 4-quark 
operators gave a total uncertainty in the hyperfine splitting of 9 MeV~\cite{Dowdall:2011wh}, 
where the error is now dominated by missing $\mathcal{O}(v^6)$ 
relativistic corrections. 
Ref.~\cite{Dowdall:2011wh} also included a calculation of the ratio of the $2S$ to $1S$ hyperfines, which is very insensitive to $c_4$, finding 0.499(42). Ref.~\cite{Meinel:2010pv} included spin-dependent $\mathcal{O}(v^6)$  corrections with $c_4 $ 
tuned from experimental results for P-wave splittings in the Upsilon system. The results were a 1S hyperfine splitting of 60.3(7.7) and a 2S to 1S hyperfine ratio of 0.403(59). 

Here we go beyond previous calculations by including radiative 
corrections to $c_4$, spin-dependent $\mathcal{O}(v^6)$  terms and four-quark 
operators which enter at $\mathcal{O}(\alpha_s^2 v^3)$. By power counting these 
terms could have a similar impact to the $\alpha_s v^4$ terms. 
These improvements mean a reduction in systematic errors 
to the level of 6.4 MeV.

We begin by describing the improvements made to the NRQCD action in Section \ref{sec:lattice}, before our results in Section \ref{sec:results} and a discussion in Section \ref{sec:conclusions}.

%
\section{Lattice calculation}
\label{sec:lattice}
%

We use a lattice NRQCD action correct through $\order{(v^4)}$ that includes discretisation corrections, spin-dependent terms through $\order{(v^6)}$ and four-quark contact interactions that enter at $\order{(\alpha_s^2 v^3)}$ \cite{Thacker:1990bm,Lepage:1992tx}. 

As described in Ref. \cite{Thacker:1990bm}, the $b$-quark velocity in bottomonium is of order $v^2 \sim 0.1$. For power-counting estimates, we assume $\alpha_s \sim 0.3$. The  $\order{(v^4)}$ action, with radiative corrections discussed below, has already been used to study $S$-, $P$- and $D$-wave splittings in Refs. \cite{Dowdall:2011wh,Daldrop:2011aa} and B-meson mass splittings in \cite{Dowdall:2012ab,Dowdall:2013tga}.
The Hamiltonian includes the leading order term $aH_0$ and sub-leading 
terms $a\delta H_{v^4}$, $a\delta H_{v^6}$ and $a\delta H_{4q}$ as:
\begin{eqnarray}
 aH_0  &=& - \frac{\Delta^{(2)}}{2 am_b}, \nonumber \\
a\delta H_{v^4}
&=& - c_1 \frac{(\Delta^{(2)})^2}{8( am_b)^3}
    + c_2 \frac{i}{8(am_b)^2}\left(\bf{\nabla}\cdot\tilde{\bf{E}}\right. -
          \left.\tilde{\bf{E}}\cdot\bf{\nabla}\right) \nonumber \\
& & - c_3 \frac{1}{8(am_b)^2} \bf{\sigma}\cdot\left(\tilde{\bf{\nabla}}\times\tilde{\bf{E}}\right. -
          \left.\tilde{\bf{E}}\times\tilde{\bf{\nabla}}\right) \nonumber \\
& & - c_4 \frac{1}{2 am_b}\,{\bf{\sigma}}\cdot\tilde{\bf{B}}  
    + c_5 \frac{\Delta^{(4)}}{24 am_b} 
- c_6 \frac{(\Delta^{(2)})^2}{16n(am_b)^2},\nn \\
a\delta H_{v^6} 
&=& - c_7 \frac{1}{8(am_b)^3} \left\{ \Delta^{(2)}, \sigma \cdot \tilde{\bf B}   \right\} \nn \\
& & - c_8 \frac{3i}{64(am_b)^4} \left\{ \Delta^{(2)}, \bf{\sigma}\cdot\left(\tilde{\bf{\nabla}}\times\tilde{\bf{E}}\right. -
          \left.\tilde{\bf{E}}\times\tilde{\bf{\nabla}}\right)   \right\} \nn \\
& & + c_9 \frac{1}{8(am_b)^3} \sigma \cdot \tilde{\bf{E}} \times \tilde{\bf{E}} 
\label{eq:deltaH}
\end{eqnarray}
$am_b$ is the bare $b$ quark mass, $\tilde{\bf{E}},\tilde{\bf{B}}$ are improved field strengths and
$\Delta^{(2)},\Delta^{(4)},\bf{\nabla},\bf{\tilde \nabla} $ are lattice derivatives that are described in \cite{Gray:2005ur}.
The $c_i$ are the Wilson coefficients of the effective action; the terms are normalised so that they have the expansion $c_i = 1 + \alpha_s c_i^{(1)}$. The matching coefficients for the kinetic terms $c_1,c_5,c_6$ and the spin-magnetic coupling $c_4$ are given for the $\order(v^4)$ action in Ref. \cite{Dowdall:2011wh}. The matching of the Darwin, $c_2$, and spin-magnetic terms are described in detail for both the $\order(v^4)$ and $\order(v^6)$ actions in Ref. \cite{Hammant:2013sca}. In principle the $\order(v^6)$ terms will have an effect on the renormalisation of the kinetic terms but since these have a negligible effect on the hyperfine splittings (compare with Refs. \cite{Gray:2005ur,Dowdall:2011wh}) we neglect this small effect. These terms also affect the error on the tuning of the $b$-quark and the determination of the lattice spacing. In practice we find that the coefficient of the $\sigma.{\bf B}$ term also changed very little when adding the $\order(v^6)$ terms. 
The parameters used for the NRQCD valence quarks are given in Table \ref{tab:params}.

The four-quark interactions relevant for the hyperfine splitting are
\begin{eqnarray}
\label{eq:4qlagrangian}
 \mathcal{L}_{4q} &=&  \frac{d_1\alpha_s^2}{(am_b)^2} \psi^\dagger \chi^* \chi^T \psi 
 +  \frac{d_2\alpha_s^2}{(am_b)^2} \psi^\dagger \sigma^i \chi^* \chi^T \sigma^i \psi,
\end{eqnarray}
where $\psi,\chi$ are the quark and antiquark respectively. These operators cannot be included directly in the Hamiltonian, since they involve both the quark and antiquark, but they can be implemented stochastically with a Hubbard-Stratonovich transformation \cite{Lepage:1992tx}. The quarks are propagated in an auxiliary complex Gaussian noise field $\eta$ with two colour and two spin indices using the following Lagrangian
\be
 \mathcal{L}_{4q} =  z \psi^\dagger \eta \psi +  z^* \chi^T \eta^\dagger \chi^* , \ \ \ 
z= -\frac{d_1^{\frac{1}{2}}  \alpha_s}{am_b} ,
\ee
with $\eta$ normalised such that $\langle \eta \eta^\dagger \rangle = 1$.
This requires that the quark Hamiltonian includes a term $a\delta H_{4q} = z\eta$, and similarly for the anti-quark.
By solving the equations of motion, one can show that this is equivalent to the first term in Eq. (\ref{eq:4qlagrangian}). A similar method can be used for the second term.

The spin-dependent contribution to the coefficients of the four-quark operators was determined in \cite{Hammant:2011bt,Hammant:2013sca} and $d_1$ includes a contribution $-2\alpha_s^2 (2-\ln 2)/9$ from $\bar b b$ annihilation. We use $\alpha_V$ at $q=\pi/a$.  The coefficients used on each ensemble are given in Table \ref{tab:params}.

We employ three ensembles of gluon configurations at different lattice spacings but with the same light quark masses, see Table \ref{tab:gaugeparams}. We demonstrated in \cite{Dowdall:2011wh} that the light quark mass dependence of hyperfine splittings is much smaller than our other systematic errors.
The bare lattice $b$-quark mass was tuned using the spin averaged kinetic mass. As expected, the tuned values given in Ref. \cite{Dowdall:2011wh} did not change within errors when the $\order(v^6)$ terms were added. We use the retuned values which are slightly different to those in Ref. \cite{Dowdall:2011wh}.
To reduce statistical errors we used $U(1)$ random wall sources on 16 time slices and used five smearing combinations as described in Ref. \cite{Dowdall:2011wh}. All correlators on the same configuration are binned. Correlators are fit using a simultaneous multi-exponential Bayesian fit \cite{gplbayes}; however the vector and pseudo-scalar states were fit separately. 
Autocorrelations and finite-volume effects are negligible for low-lying bottomonium states \cite{Dowdall:2011wh}.

\begin{table}
\caption{ 
\label{tab:params}
Parameters used for the valence quarks. $am_b$ is the bare $b$ quark mass in lattice units, and $u_{0L}$ is the Landau link. 
The $c_i, d_i$ are coefficients of terms in the NRQCD action [see Eq. (\ref{eq:deltaH})]. $c_4, d_1$ and $d_2$ use $\alpha_s$ in the V-scheme at scale $\pi/a$. The other coefficients use different scales as discussed in \cite{Dowdall:2011wh}.
}
\begin{ruledtabular}
\begin{tabular}{lllllllllll}
Set & $am_b$ & $u_{0L}$ & $c_1,c_6$ & $c_5$ & $c_4$ & $d_1\alpha_s^2$ & $d_2\alpha_s^2$ & $\alpha_s(\pi/a)$\\
\hline
1 & 3.31 & 0.8195    & 1.36 & 1.21 & 1.23 & -0.1021 & 0.0306 & 0.275 \\
2 & 2.73 & 0.834      & 1.31 & 1.16 & 1.19 & -0.058  & 0.016  & 0.255 \\ 
3 & 1.95 & 0.8525     & 1.21 & 1.12 & 1.18 & -0.026  & 0.006  & 0.225 \\ 
\end{tabular}
\end{ruledtabular}
\end{table}

\begin{table}
\caption{
Parameters of the gauge configurations used. $\beta$ is the Yang-Mills coupling, $a_{\Upsilon}$ is the lattice scale in fm determined using the $\Upsilon(2S-2S)$ splitting. $am_q$ are the sea quark masses, $L,T$ are the spatial and temporal extents of the lattice and $n_{{\rm cfg}}$ is the size of the ensemble.
}
\label{tab:gaugeparams}
\begin{ruledtabular}
\begin{tabular}{lllllllll}
Set & $\beta$ & $a_{\Upsilon}$ (fm) 	& $am_{l}$ & $am_{s}$ & $am_c$ & $L \times T$ & $n_{{\rm cfg}}$  \\
\hline
1 & 5.8 & 0.1474(5)(14)(2)  & 0.013   & 0.065  & 0.838 & 16$\times$48 & 1020 \\
2 & 6.0 & 0.1219(2)(9)(2)   & 0.0102  & 0.0509 & 0.635 & 24$\times$64 & 1052 \\
3 & 6.3 & 0.0884(3)(5)(1)   & 0.0074  & 0.037  & 0.440 & 32$\times$96 & 1008 \\
\end{tabular}
\end{ruledtabular}
\end{table}

%
\section{Results}
\label{sec:results}
%

\begin{table}
\caption{ 
\label{tab:results}
Fit results in lattice units for the $1S$ and $2S$ energies of the $\Upsilon$ and $\eta_b$ for the $\order(v^6)$ action
with and without four-quark interactions. 
}
\begin{ruledtabular}
\begin{tabular}{lllll}
Set & $aE_\Upsilon(1S)$ & $aE_\Upsilon(2S)$ & $aE_{\eta_b}(1S)$ & $aE_{\eta_b}(1S)$ \\
\hline
\multicolumn{5}{l}{$\order(v^6)$ action } \\
1 &  0.27547(6) & 0.6917(14) & 0.22950(5) & 0.6748(12) \\
2 &  0.28786(3) & 0.6332(9)  & 0.25148(3) & 0.6180(6)  \\
3 &  0.30269(3) & 0.5573(10) & 0.27484(2) & 0.5461(9)  \\
\multicolumn{5}{l}{$\order(v^6)$ action + four-quark interactions }  \\
1 & 0.27699(6)  & 0.6997(13) & 0.22460(5) & 0.6755(13) \\
2 & 0.28850(3)  & 0.6340(10) & 0.24897(3) & 0.6166(9) \\
3 & 0.30271(2)  & 0.5564(6)  & 0.27378(2) & 0.5437(5) \\
\end{tabular}
\end{ruledtabular}
\end{table}

Our results are the hyperfine splitting $\Delta_{\rm{hyp}}(1S) = M_{\Upsilon}(1S) - M_{\eta_b}(1S)$ and the ratio of hyperfine splittings $R_H = \Delta_{\rm{hyp}}(2S)/\Delta_{\rm{hyp}}(1S)$. The ratio can be calculated much more accurately than the separate splittings since the dominant NRQCD systematics cancel. 
We find that the effect of the four-quark operators on the hyperfine splittings is within $2$ MeV (or $4\%$) of the expected shift calculated in perturbation theory. This is given by $6\alpha_V^2(\pi/a)(d_1-d_2)|\psi(0)|^2/m_b^2$,  where $\psi(0)$ is the wavefunction at the origin obtained from the spin-averaged amplitude of the ground state at each lattice spacing.

The fitted $\Upsilon$ and $\eta_b$ energies are given in Table \ref{tab:results} for the $v^6$ action with and without the four-quark operators.  To extract a physical result $f_{\rm{phys}}$, and determine the uncertainty due to scale dependence, we fit results from all three lattice spacings to the form
\begin{multline}
\label{eq:fitxa}
f(a^2,am_b) 
= 
f_{\mathrm{phys}}\\
\times \left[ 1+\sum_{j=1,2}k_j(a\Lambda)^{2j}(1 +k_{jb}\delta x_m + k_{jbb}(\delta x_m)^2)  \right] .
\end{multline}
The lattice spacing dependence is set by a scale $\Lambda=500$ MeV, and 
$\delta x_m = (am_b - 2.7)/1.5$ allows for mild dependence on the effective 
theory cutoff $am_b$. We take Gaussian priors of mean $0$ and width $1$ on all 
the coefficients in the fit to the 1S hyperfine splitting. 
We tighten the prior on $k_{1}$ to 0.0(3) in fitting $R_H$ since 
the action includes radiatively improved $a^2$ lattice spacing corrections and we also expect 
cancellation of discretisation errors in the ratio. 
We have tested that our results are not sensitive to the fit form or the priors.

A number of systematic errors must be accounted for in our calculation. Some of these depend on the lattice scale and are included in the fit, while others are estimated at the end. The dominant error comes from missing radiative corrections to $c_4$ since the hyperfine splitting is proportional to $c_4^2$ at leading order. We multiply each raw data point by a Gaussian error with unit mean and width $2\alpha_v^2(\pi/a)$ which assumes an $\alpha_s^2$ coefficient of order one. We enforce that this component of the error is correlated between each lattice spacing, i.e. that considering this error in isolation the correlation coefficient between data on two lattice spacings is $\rho = 1$. In practice, this takes the value on the fine lattice as the $\alpha_s^2 v^4$ error. We allow for the statistical error in the determination of $c_4^{(1)}$ \cite{Hammant:2013sca} coming from the Vegas integration by  adding an error of  $ \delta c_4^{(1)} \alpha_V(\pi/a)|\psi(0)|^2/m_b^2$, which comes from the 1-loop perturbative expression for the hyperfine splitting and is a sufficient approximation for estimating this component of the error.
Higher order corrections to the four-quark operators are also significant, and depend on $|\psi(0)|^2$. We allow a correlated additive error of the form $6\alpha_V^3(\pi/a)|\psi(0)|^2/m_b^2$ with a coefficient $(\pm1 \pm \ln(am_b))$. The four-quark error is applied to both splittings in the ratio $R_H$ but the systematic from $c_4$ cancels.
We allow an additional (small) error from uncertainty in the $b$-quark mass, using the systematic errors from Ref. \cite{Dowdall:2011wh} and assuming the leading order dependence $\propto 1/m_b^2$. The error in $am_b$ from scale setting is included by taking twice the lattice spacing error when converting $\Delta_{\rm{hyp}}(1S)$ to GeV.

The dominant uncertainty in the hyperfine splitting is from corrections to the $v^4$ term $c_4$ but we also include estimates of the uncertainty from higher order terms in the action since these are now relevant. The first is for radiative corrections to $c_7$ which is the term at order $v^6$ that contributes to the hyperfine splitting. This uncertainty is not as straightforward as for the $c_4$ term since it is subleading, but we can allow a naive power-counting estimate of the error. The hyperfine splitting is an $\mathcal{O}(v^4)$ effect and these terms are subleading by an additional factor of $v^2$ so we take an error of $2v^2 \alpha_s(\pi/a)$. As above, we make this component of the error correlated between all the lattice spacings.  We verified that this is reasonable by running with a different value of $c_7=1.25$ on the $0.09$ fm ensemble. This shifts the hyperfine splitting down by 2 MeV, which is within the error. Finally, we take a multiplicative $1\%$ error on the final answer to allow for missing $v^8$ terms in the action. 

As an aside, we verified that corrections to $c_7=1$ have the desired effect on the kinetic mass. As discussed in Ref.~\cite{Dowdall:2011wh}, the hyperfine splitting calculated from the kinetic masses of the $\Upsilon$ and $\eta_b$ using a $v^4$ action is incorrect. To get the correct hyperfine splitting from kinetic masses, relativistic corrections to the $\sigma \cdot {\bf{B}}$ term must be present in the Hamiltonian; for a detailed discussion see Ref.~\cite{Dowdall:2011wh}. For the pure $v^6$ action (with $c_7=1$), we obtained $29(23)_{\rm stat}$ MeV for the kinetic hyperfine splitting 
and with $c_7=1.25$ we obtained $52(17)_{\rm stat}$ MeV. This verifies that shifts to  $c_7$ of the size of radiative corrections are all that is needed to obtain the correct kinetic hyperfine splitting.

Effects from electromagnetism can be estimated from a potential model. The dominant effect of the Coulomb interaction between $b$ and $\bar{b}$ (missing from our calculation) was estimated in Ref. \cite{Gregory:EM} to give an upward shift of $1.6$ MeV to both $\Upsilon$ and $\eta_b$ and therefore with no impact on the hyperfine splitting. Relativistic corrections to this might then be expected to give an effect on the hyperfine splitting of approximately $10\%$ of this, or $0.2$ MeV ($0.3\%$).

Our final results are
\begin{eqnarray}
\Delta_{\rm{hyp}}(1S) &=&  60.0(6.4) {\rm\  MeV}             \nn \\
 \Delta_{\rm{hyp}}(2S)/\Delta_{\rm{hyp}}(1S) &=& 0.445(28).
\end{eqnarray}

\begin{figure}
\includegraphics[width=\hsize]{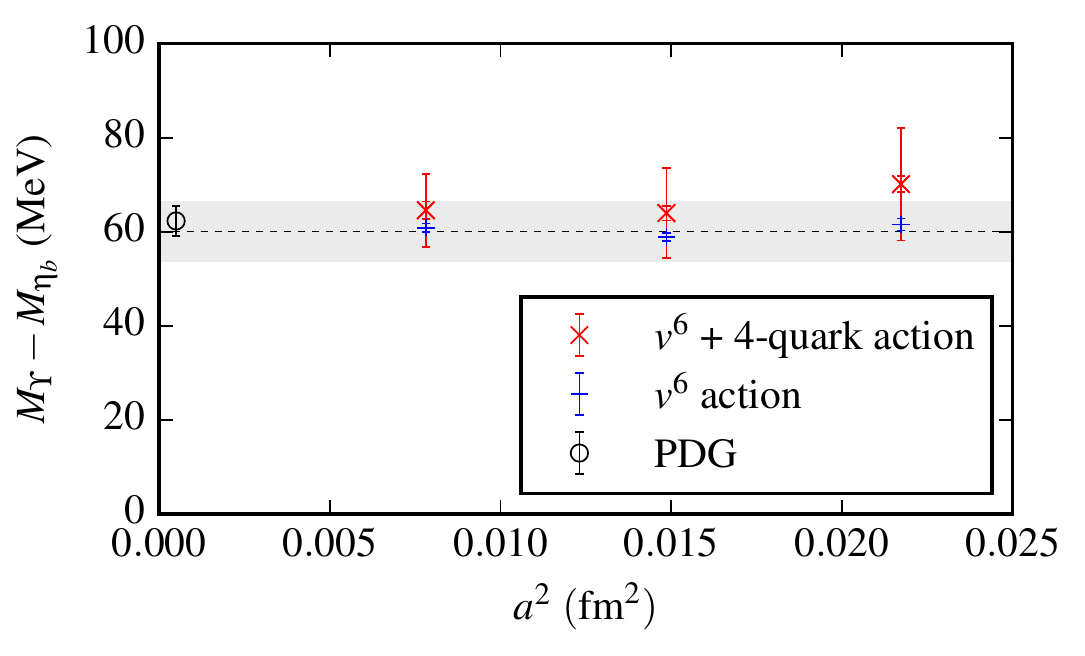}
\caption{\label{fig:hyp1s} (color online). The fit result for the $1S$ hyperfine 
splitting from the $v^6$ action with spin-dependent four-quark operators.
The two error bars are: statistics and scale setting for the smaller; the larger band includes correlated errors from missing radiative corrections to $c_4,d_1,d_2$ and 
quark mass tuning errors.  
The grey band is the final physical result and includes all systematic errors. 
Also shown are the pure $v^6$ results (statistical errors only), which are not included in the fit, and the PDG average \cite{pdg}.}
\end{figure}

\begin{figure}[t!]
\includegraphics[width=0.99\hsize]{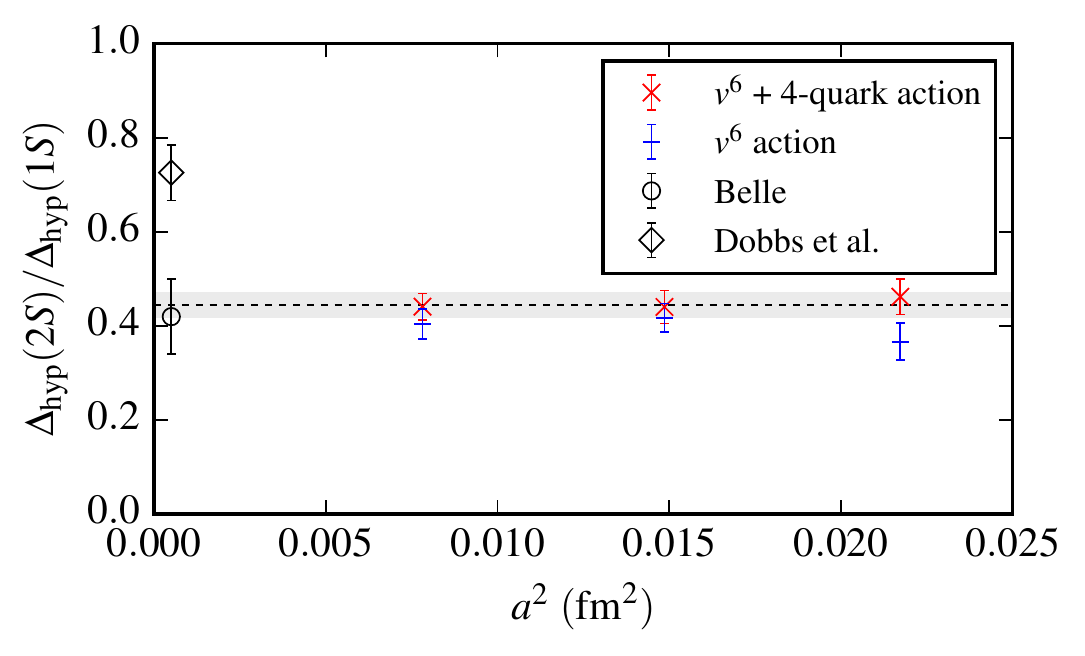}
\caption{\label{fig:hyp2s} (color online). The fit result for the hyperfine ratio from the $v^6$ action with spin-dependent four-quark operators. The grey band is the final physical result and includes all systematic errors. Also shown are the pure $v^6$ results, which are not included in the fit, and the results from Belle and Dobbs et al. \cite{belleetab,cleoetab2s}.}
\end{figure}

The fit results are plotted in Figs. \ref{fig:hyp1s}, \ref{fig:hyp2s} and a full error budget is given in Table \ref{tab:errorbudget}. 
The error on $\Delta_{\rm{hyp}}(1S)$ is dominated by uncertainties in $c_4$. Reducing the systematic error would require a difficult 2-loop calculation.
The $\alpha_s^3$ corrections to the four-quark operators are also significant; again improving these further would be difficult.
Missing higher-order operators are no longer a significant source of error.
Statistical errors dominate the uncertainty in the ratio which could in principle be improved. Sea quark mass dependence in the hyperfine splitting was found to be much less than other errors in Ref. \cite{Dowdall:2011wh} so we neglect it in $\Delta_{\rm hyp}(1S)$. For the ratio, we also found no systematic light-quark mass dependence in Ref. \cite{Dowdall:2011wh}, but statistical fluctuations between different ensembles accounted for 3.5\% of the error. We apply this additional error to our result for the ratio.

\begin{table}
\caption{
\label{tab:errorbudget}
Full error budget for the $1S$ hyperfine splitting 
and the $2S$ to $1S$ ratio. All errors are in percent. }
\begin{ruledtabular}
\begin{tabular}{lcccc}
Error \%			& $\Delta_{\rm{hyp}}(1S)$ & $R_H$     \\
\hline
Stats/fitting    			 & $~~$0.2	&	$~~$4.1	  \\
Uncertainty in $a$    			 & $~~$2.2	&	$~~$0.0	  \\
scale dependence 	        	 & $~~$1.3	&	$~~$2.5	  \\
NRQCD $am_b$ dependence       		 & $~~$3.6	&	$<$0.0	  \\
NRQCD radiative $\alpha_s v^6$		 & $~~$3.7	&	$~~$0.0	  \\
NRQCD radiative $\alpha_s^2 v^4$ in $c_4$& $~~$7.1	&	$<$0.1	  \\
Statistical error in $c_4^{(1)}$         & $~~$3.2  &           $~~$1.4      \\
NRQCD relativistic spin $v^8$		 & $~~$1.0	&	$~~$0.5	  \\
NRQCD radiative four-quark $\alpha_s^3 v^3$ & $~~$3.0	&	$~~$1.3	  \\
$m_b$ tuning				 & $~~$0.7	& 	$<$0.1	  \\
$m_{\rm l, sea}$ dependence              & $<$0.1    &          $~~$3.5       \\
EM effects                               & $~~$0.3       &      $<$0.1       \\
\hline
Total 	 	              		& $~~$10.8   	  &     $~~$6.3    \\
\end{tabular}
\end{ruledtabular}
\end{table}


%
\section{Conclusions}
\label{sec:conclusions}
%

\begin{figure}
\vspace{2mm}
\includegraphics[width=\hsize]{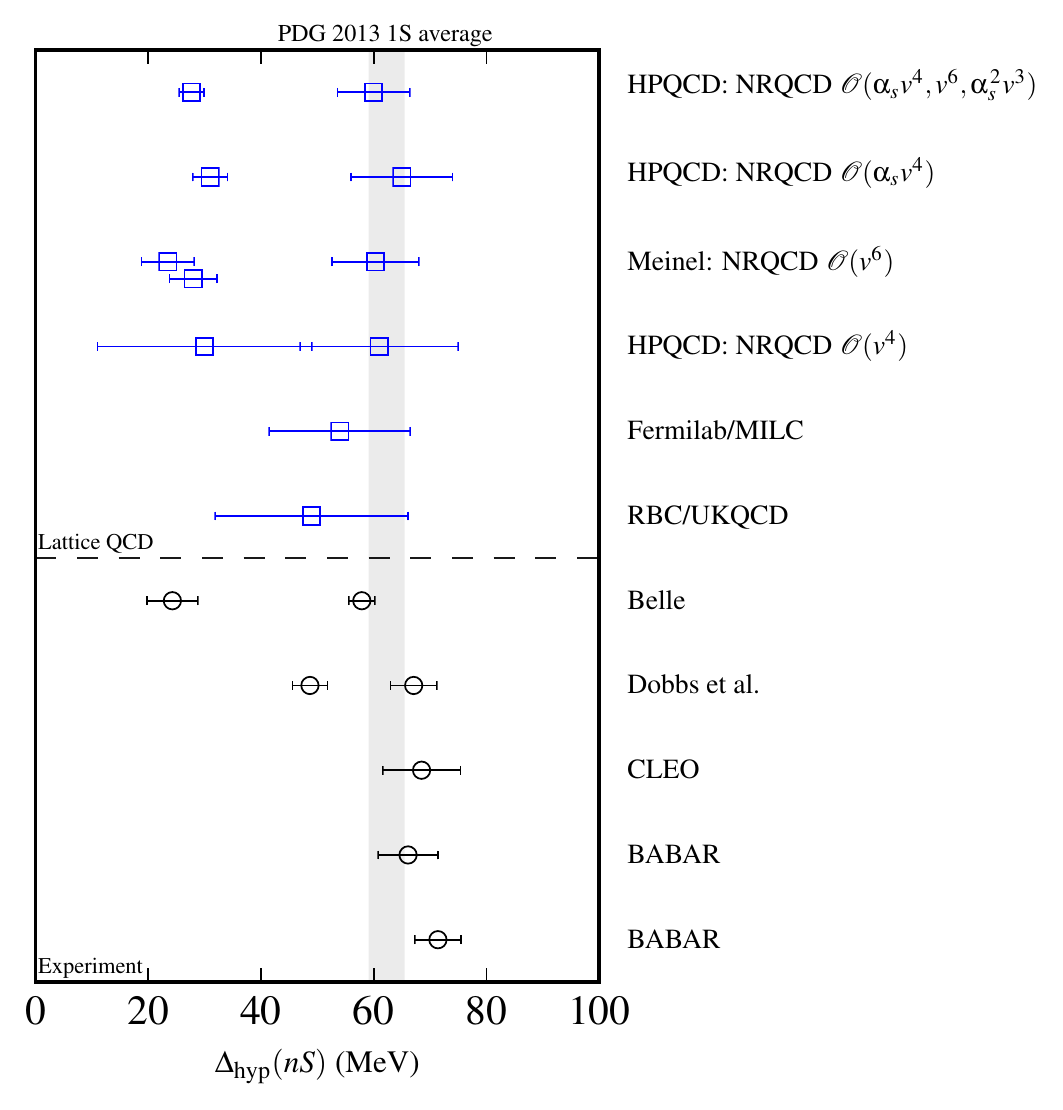}
\caption{\label{fig:hypcomparison} (color online). Comparison of our result (top) with other lattice calculations \cite{Aoki:2012xaa,Meinel:2010pv,Burch:2009az,Gray:2005ur} and experimental \cite{pdg,belleetab,cleoetab2s,Bonvicini:2009hs,Aubert:2009pz,Aubert:2008ba}
results for the hyperfine splittings. 
The $2S$ hyperfine splitting is on the left and the $1S$ on the right. 
$\Delta_{\rm hyp}(1S)$ for the earlier HPQCD $\mathcal{O}(\alpha_sv^4)$ result 
has been adjusted to use the correct $d_1, d_2$ coefficients for that case 
and the $\Delta_{\rm hyp}(2S)$ value plotted uses the ratio of 2S to 1S splittings 
given in that paper multiplied by the 2013 PDG average for $\Delta_{\rm hyp}(1S)$. 
Meinel quotes two results for $\Delta_{\rm hyp}(2S)$ normalising the result using either the $1S$ hyperfine or the $1P$ tensor splitting. The results are taken as given in Ref. \cite{Meinel:2010pv}. }
\end{figure}

The bottomonium spectrum continues to provide a rich environment for increasingly precise tests of QCD in the low energy regime. Excited states are still being discovered by experiments such as Belle, CLEO and ATLAS \cite{belleetab,bellehb,cleoetab2s,Aad:2011ih} and lattice QCD calculations are now able to accurately calculate most of the low-lying states. 
We have given an improved determination of the hyperfine splittings using 
nonrelativistic QCD correct through $\order(\alpha_sv^4,v^6, \alpha_s^2 v^3)$. 
Our result for the 1S hyperfine splitting is
\begin{equation}
\Delta_{\rm{hyp}}(1S) =  60.0(6.4) {\rm\  MeV} .  
\end{equation}
This is the most accurate calculation to date and supersedes all of our earlier calculations. 
The result agrees with, but is more accurate than other results in full lattice QCD, those of Meinel \cite{Meinel:2010pv}, Fermilab/MILC \cite{Burch:2009az} and RBC/UKQCD \cite{Aoki:2012xaa}. 
It is also consistent with the PDG average for the experimental 1S hyperfine splitting~\cite{pdg}
as well as with the most recent and accurate result from Belle~\cite{belleetab}.

Our result for the ratio $R_H$ of 2S to 1S hyperfine splittings is in excellent 
agreement with the Belle result but disagrees significantly with Dobbs et al. 
For  discussion of the discrepancy in the experimental values, see Ref. \cite{mizukreview}. 
Using the current PDG average of 62.3(3.2) MeV for the 1S hyperfine gives a 
2S hyperfine splitting of 
\begin{equation}
 \Delta_{\rm hyp}(2S) = 27.7(1.7)_{\rm latt}(1.4)_{\rm exp},
\end{equation}
where the error has been divided into components from this lattice calculation and the experimental result. Using our lattice result to normalise the value gives a consistent result. 

A comparison of the existing lattice and experimental results is 
shown in Fig. \ref{fig:hypcomparison}. Our new results are shown as 
the top row. We have adjusted the value of our earlier 
1S hyperfine splitting from~\cite{Dowdall:2011wh} since this included an 
estimate of the effect of 4-quark operator terms using incorrect coefficients 
for $d_1$ and $d_2$. We estimate the correction using a potential model 
analysis, as done in~\cite{Dowdall:2011wh}, but now using the correct $d_1$ 
and $d_2$ for that action. This gives 65(9) MeV.  
We also determine a new 2S hyperfine splitting value using the ratio 
of 2S to 1S splittings and the updated value for the experimental 1S splitting. 
Good agreement between all of the lattice QCD results is evident. 

Further improvements to lattice calculations of the hyperfine splitting may require relativistic actions. A calculation of the hyperfine splitting using HISQ $b$-quarks is in progress.

%
\section*{Acknowledgements}
\label{sec:acknowledgements}
%
We are grateful to the MILC collaboration for the use of their 
gauge configurations. 
This work was funded by STFC.
CTHD is supported by the Royal Society and the Wolfson Foundation.
The results described here were obtained using the Darwin Supercomputer 
of the University of Cambridge High Performance 
Computing Service as part of the DiRAC facility jointly
funded by STFC, the Large Facilities Capital Fund of BIS 
and the Universities of Cambridge and Glasgow.


\bibliography{bottomonium_bib}

\end{document}